\documentclass{acm_proc_article-sp}
\usepackage[
  pass,
]{geometry}

\newfont{\mycrnotice}{ptmr8t at 7pt}
\newfont{\myconfname}{ptmri8t at 7pt}
\let\crnotice\mycrnotice%
\let\confname\myconfname%
\usepackage{url}
\begin{document}

\title{A directive based hybrid Met Office NERC Cloud model}
\numberofauthors{6}
\author{
\alignauthor Nick Brown\\
\affaddr{EPCC, James Clerk Maxwell Building, Peter Guthrie Tait Road, Edinburgh}\\
\email{nick.brown@ed.ac.uk}
\alignauthor Angus Lepper\\
\affaddr{EPCC, James Clerk Maxwell Building, Peter Guthrie Tait Road, Edinburgh}
\alignauthor Michele Weiland\\
\affaddr{EPCC, James Clerk Maxwell Building, Peter Guthrie Tait Road, Edinburgh}
\and
\alignauthor Adrian Hill\\
\affaddr{UK Met Office, FitzRoy Road, Exeter, Devon}
\alignauthor Ben Shipway\\
\affaddr{UK Met Office, FitzRoy Road, Exeter, Devon}
\alignauthor Chris Maynard\\
\affaddr{UK Met Office, FitzRoy Road, Exeter, Devon}
}

\maketitle
\begin{abstract}
Large Eddy Simulation is a critical modelling tool for the investigation of atmospheric flows, turbulence and cloud microphysics. The models used by the UK atmospheric research community are homogeneous and the latest model, MONC, is designed to run on substantial HPC systems with very high CPU core counts. In order to future proof these codes it is worth investigating other technologies and architectures which might support the communities running their codes at the exa-scale. 

In this paper we present a hybrid version of MONC, where the most computationally intensive aspect is offloaded to the GPU while the rest of the functionality runs concurrently on the CPU. Developed using the directive driven OpenACC, we consider the suitability and maturity of this technology to modern Fortran scientific codes as well general software engineering techniques which aid this type of porting work. The performance of our hybrid model at scale is compared against the CPU version before considering other tuning options and making a comparison between the energy usage of the homo- and hetero-geneous versions. The result of this work is a promising hybrid model that shows performance benefits of our approach when the GPU has a significant computational workload which can not only be applied to the MONC model but also other weather and climate simulations in use by the community.
\end{abstract}

\keywords{OpenACC, MONC, LEM, GPU, Large Eddy Simulation, Hybrid}

\section{Introduction}
Large Eddy Simulation (LES) is a computational fluid dynamics technique used to efficiently simulate and study turbulent flows. In atmospheric science, LES is often coupled to cloud microphysics and radiative transfer schemes, to create a high resolution modelling framework that is employed to understand the physics of these turbulent flows and further develop and test physical parametrisations and assumptions used in numerical weather and climate prediction. The Met Office NERC Cloud model (MONC) \cite{easc} is an LES that is being developed to replace an existing model called the Large Eddy Model (LEM) \cite{lem}. The LEM has been an instrumental tool, used by the weather and climate communities, since the 1980s for activities such as development and testing of the Met Office Unified Model (UM) boundary layer scheme \cite{lock1998}\cite{lock2000}, convection scheme \cite{petch2001}\cite{petch2006} and cloud microphysics \cite{abel2007}\cite{hill2014}. Given the solid scientific basis of the LEM, as established by many inter-comparison studies including \cite{klein2009intercomparison}\cite{ovchinnikov2014intercomparison}\cite{vanzanten2011controls}\cite{fridlind2012comparison}, the LEM is currently the principal LES employed by the UK Met Office and UK academia. 

In order to further the state of the art, scientists wish to model at a greater resolution and/or near real time which requires large amounts of computational resource. The use of modern HPC machines is crucial, however the LEM struggles to scale beyond a few hundred cores. As such we are using modern parallelisation techniques in developing MONC so that the communities are able to perform their next generation of science. One of the important aspects of this new model is the exploration of, and flexibility to be able to use, different approaches and architectures for parallelisation. To this end MONC has been designed in a modular fashion so that it is trivial to add or remove functionality without impacting other areas of the model. At the time of writing MONC is nearing an initial release and it is the hope of the development team that this community code will replace the LEM and become the de-facto LES used within the UK weather and climate communities.

Programming accelerators via directives is an \linebreak[0]{} attractive proposition where, using technologies such as OpenACC, the programmer can often keep the structure of their code unchanged and simply add directives, often focused around loops and data. However, arguably this technology has not yet reached full maturity and whilst support in some areas is good, the directives community requires application developers to port non-trivial codes using these approaches in order to understand what additional features would be useful and to stress test supporting compilers. 

By its very nature MONC is computationally intensive and in this paper we discuss and evaluate the \linebreak[0]{} experimental OpenACC port of these aspects of the model onto GPUs. Our approach is such that MONC runs in a hybrid fashion, where the GPUs and CPUs are concurrently working on different parts of the model in order to make the most efficient use of these technologies. Section 2 lays the foundation of this work and begins by explaining the architecture of MONC in detail. This section then discusses how a simulation proceeds and identifies related work done by the community in porting other weather models to GPUs and their experiences of this. Section 3 identifies the approach we have taken in making use of GPUs to turn MONC into a heterogeneous code, along with fundamental choices made about issues such as data movement and synchronisation which are important for performance reasons. Additionally there is a discussion about our experiences in terms of the suitability of OpenACC with modern Fortran. Section 4 presents the performance results of our work running at scale on Piz Daint. Optimisations to further tune the hybrid model are also explored, and an analysis of the impact of heterogeneity on energy usage is conducted.

\section{Background}
\subsection{MONC}
\label{moncbg}
MONC has been developed in Fortran 2003 and uses MPI for parallelism. The standard homogeneous version of this code is designed to be run on many thousands of cores and has demonstrated good performance and scalability on up to 32768 cores \cite{easc}. This model has been designed around pluggable components where the majority of the code complexity, including all of the science and parallelisation, are contained within these independent units. Components are managed by a registry and at run-time the user selects, via a configuration file, which components to enable or disable. The majority of a component's functionality is contained within optional callback procedures, which are called by the model at three stages: upon initialization, for each timestep and upon model completion.  There are no global variables in MONC, but instead a user derived type is used to represent the current state of the model and this is passed into each callback which itself may modify the model state. Using this approach means that the model's current state is represented in a structured manner and there is a single point of truth about the model's status at any point in time. Isolating these components significantly limits interaction between them and as such results in an easy to understand architecture where aspects can easily be added, removed or replaced without having to worry about unintended side effects. 

Via our component based architecture it is trivial to add significantly distinct functionality, for instance accelerator parallelisation via OpenACC, without it impacting on other parts of the model. Each component can be given bespoke compiler commands and directives during build time. Once components are compiled into the code then they are configured via a configuration file which specifies aspects such as whether they are enabled, when they should run and other component specific options. There is no problem including both the existing CPU component and new GPU component within the same executable. Which one is run is determined at runtime by the user configuration file and this flexibility allows for very easy correctness testing and performance analysis.

\begin{figure}
\begin{center}
\includegraphics[scale=0.5]{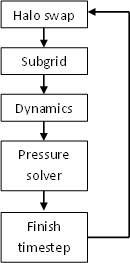}
\end{center}
\caption{MONC timestep structure}
\label{fig:tsstructure}
\end{figure}

Like many LES models the simulation proceeds in timesteps, gradually increasing the simulation time on each iteration until it reaches a predefined termination time. The model works on prognostic fields, \emph{u}, \emph{v} and \emph{w} for wind in the \emph{x}, \emph{y} and \emph{z} dimensions, \emph{$\theta$} for the temperature and any number of \emph{q} fields which represent aspects such as moisture and tracers. Figure \ref{fig:tsstructure} illustrates high level groups, each of which contains any number of components, that makes up the structure of a single timestep. Each of these groups must execute sequentially so that one group can not start until the previous group has completed. Initially all prognostic fields are halo swapped between neighbouring processes and then the sub-grid group of components are called to determine model parameterisations. The dynamics group of components, often referred to as the dynamical core, performs Computational Fluid Dynamics (CFD) in order to solve modified Navier-Stokes equations \cite{lem-science} which is followed by the pressure solver, using either an FFT or iterative approach to solve the Poisson equation. The timestep then concludes with some miscellaneous functionality such as checking for model termination.  The majority of the runtime is spent in the dynamical core (dynamics group) and in particular the advection scheme is the single longest running component. A number of different advection schemes are provided as components, and these require all the model's prognostic fields in order to complete their computation. 

\subsection{Related work}
There are a variety of climate and weather models used by the communities and GPU ports of some of these exist. GALES \cite{gales} is a GPU accelerated LES, implemented in CUDA and works at mixed-precision. For optimisation the implementation is flexible enough to support double, single or mixed precision floating point, where the later aims to run the majority of the model in single precision apart from some aspects, such as the Poisson solver, which require double precision. With some minor exceptions, the entirety of the computation is performed on the GPU, with the CPUs servicing the GPU and likely sitting idle for much of the computation which would be a waste of the modern heterogeneous machines. It is also not clear from \cite{gales} how their model would parallelise over multiple nodes and GPUs which is required in order to tackle the scale of problems that scientists wish to currently investigate.

The Non-hydrostatic Icosahedral Model (NIM) is a weather prediction model developed at NOAA's Earth System Research Laboratory which has been ported to GPUs and the MIC architecture \cite{nim2014}. The initial port was done using F2C-ACC directives and then the code was updated to also support OpenACC. This project has ported the dynamical core of NIM onto GPUs and \cite{nim2014} discusses in detail the optimisations performed to reach the level of performance that they have met. As illustrated in section \ref{moncbg}, a MONC timestep is made up of a number of functionality groups and each group must run sequentially, i.e. the results from one group must be known before proceeding to the next group. Therefore, it would not be efficient for us to port the entirety of the MONC dynamical core onto the GPU as the CPU would sit idle whilst the GPU is executing this aspect of the model. \cite{nim2010} discusses using the GPU version of NIM over massively parallel machine, but it is currently unclear how this work is going or its scalability.   

The High-Order Method Modeling Environment (HOMME) dynamical core was ported to GPUs in \cite{homme}. The team concentrated on the advection of tracer fields (a subset of Q fields in MONC) and found that modelling these on GPUs afforded a good performance increase on large scale architectures. However they use synchronous data transfer and execution, which means the CPU must wait for completion and, whilst kernels do work on common data, no data sharing has been implemented meaning that an excessive amount of data must be transferred to and from the device.  

COSMO is an atmospheric model used for weather prediction by a variety of European organisations and it has recently been adapted to run on GPU systems \cite{fuhrer2014towards}. They have adopted two approaches, an entire rewrite of the dynamical core and a directive based approach for other parts. The dynamical core, which takes up around 60\% of the overall runtime and is modified infrequency \cite{cosmocug}, was ported to a domain specific language, STELLA \cite{stella}, to separate out the atmospheric model from the architecture specific implementation so that the user code can run on any machine architecture transparently. Other parts of COSMO, which are less computationally intensive and modified by a wider number of people have been ported into OpenACC meaning that the community can keep the same code base for these parts without having to learn new languages. The dual approach adopted here makes sense, although for us MONC is a community code where casual users wish to modify and extend all aspects of the model. For users of MONC it is not realistic to expect them to pick up a new programming language and the requirement for familiarity is why we developed the model in Fortran in the first place. 

The hybrid set-up of COSMO experiments in \cite{fuhrer2014towards} run one MPI process per node with the GPU itself doing the majority of the computation. One might wish to place multiple MPI processes on a node to utilise the all CPU cores which could be doing useful computation on some area of the domain. The approach adopted by STELLA abstracts the user from the architecture so mixing multiple processor cores per node and the GPU should be transparent but from \cite{fuhrer2014towards} it is unclear how this and the OpenACC aspects of COSMO might achieve this in practice. It was also noted by \cite{cosmocug} that code changes, such as loop restructuring and removal of automatic arrays, were required for optimisation when using OpenACC directives. This had a severe impact upon CPU performance and as such the authors have had to maintain two separate versions of some of their code, one for CPUs and one for GPUs.

\section{OpenACC hybrid MONC}
\label{imp}
\begin{figure}
\begin{center}
\includegraphics[scale=0.5]{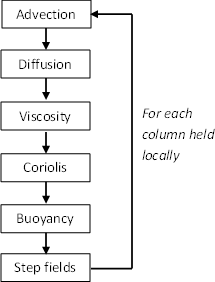}
\end{center}
\caption{Example dynamics group}
\label{fig:dynamics}
\end{figure}

The dynamical core works in a column fashion, where each component in the dynamics group is called sequentially for a specific column before the group iterates onto the next column. The exact contents of this dynamical core group depends upon the user's configuration but a common makeup is illustrated in figure \ref{fig:dynamics}. Regardless of its functionality, each component contributes the result of its specific calculation by combining (addition operator) it to an initially zero source term for the column of each prognostic field. These source terms are then integrated into the actual prognostic fields in the \emph{step fields} component. The operator which sums the contribution from each component into the source term is commutative and associative, therefore these dynamical components can be called in any order within the group without impacting correctness, as long as the \emph{step fields} integration functionality is performed last. Additionally, at the component level, there are no dependencies on source term calculations between columns. Therefore if one was to extract a component and run it on the entirety of a field (all columns) before any other components in the dynamics group then correctness would be maintained. These attributes of the dynamics group: numerous component members and the mathematical flexibility of the source term integration, are not shared with other parts of the timestep such as the pressure solver group.

\begin{figure}
\includegraphics[scale=0.36]{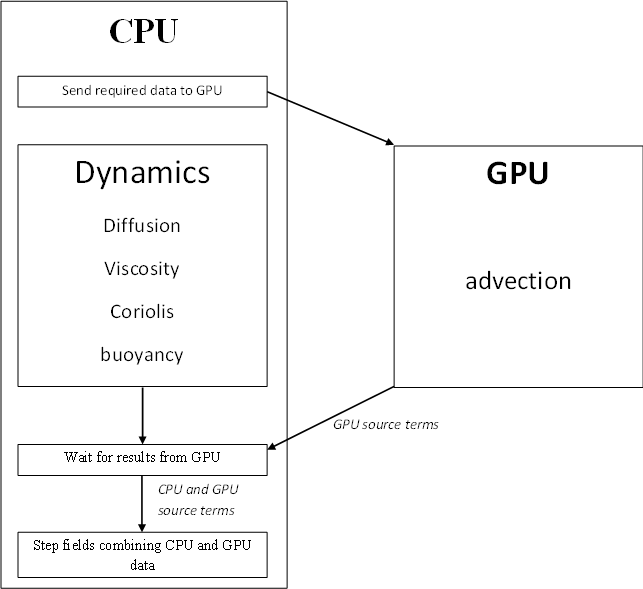}
\caption{Hybrid dynamics structure}
\label{fig:dynamics-gpu}
\end{figure}

As mentioned in section \ref{moncbg} the majority of the model runtime is spent in the advection component of the dynamics group. In order to effectively make the code heterogeneous we want to minimise the impact of the copying time onto and from the GPU device. This means running both the CPU and GPU concurrently, with them completing separate tasks which equates to running different components of the dynamics group. Therefore we have split out the advection component from this group and ported it using OpenACC to run on GPUs. Figure \ref{fig:dynamics-gpu} illustrates the hybrid execution of the model, where the CPU copies the necessary data to the device and then proceeds running the other components in the dynamics group and computing source terms for each column of the fields. At the same time, the GPU receives the required data and then uses this to run the advection kernel and compute advection source terms for the entirety of the field (each column at the same time). Once the CPU has completed the entirety of its own work it then waits for the GPU advection source terms to be made available, combines the CPU and GPU source terms together and then integrates these into the prognostic fields. 

The authors of the GPU port of HOMME \cite{homme} noted that asynchronous data transfer and execution along with data sharing between kernels would be a likely optimisation route for them to follow. Therefore in order to minimise the device data copying cost of our hybrid MONC code, copying to and from the GPU device is asynchronous which means that the CPU can initialise the data copy and as soon as this has been enqueued, proceed with its own dynamics computation work without having to wait for the copy to explicitly complete. Ideally the workload of the CPU and GPU will be roughly equal such that these complete their computation and make their results available at similar times. The results from the GPU should ideally be available to the CPU immediately after the CPU has completed its own work. Irrespective of the field being advected, the code itself requires the \emph{u}, \emph{v} and \emph{w} flow (wind) fields along with a variety of vector and scalar constants. In addition to asynchronous data transfer, another optimisation strategy adopted was to share the constants between kernels and copy these across only once (on initialisation), the \emph{u}, \emph{v} and \emph{w} fields are also shared between kernels but vary from one timestep to another and hence are copied across once, and only once, per timestep. It is only the \emph{$\theta$} (temperature) and \emph{q} (moisture and tracers) fields that are not shared between kernels and these are copied in explicitly if the user has configured the GPU to advect these quantities. 

There are a number of existing accelerator programming technologies, but OpenACC was chosen due to the ease of integration with Fortran and its maturity over other directive based accelerator approaches such as OpenMP 4.0. It was also felt that a directive based approach is preferable because the structure of the code can, in many cases, remain unchanged. Using directives the programmer is free to concentrate on higher level considerations, such as data movement and kernel placement, with the compiler dealing with the lower level, tricky and uninteresting aspects such as memory allocation and physical data movement. In modern heterogeneous supercomputers, a node typically contains many cores linked to one GPU, therefore a technology which transparently allows each core to independently launch kernels on the same GPU is important. Other technologies, such as OpenCL or CUDA were considered, but there is no port of OpenCL to Fortran and therefore this would have required wrapping the C API and making it available via Fortran's native C interface. This approach would not only have resulted in a large amount of boiler plate code but would also be a source of complexity and bugs when going between C and Fortran's different allocation ordering for arrays. CUDA was a more attractive option than OpenCL, not least because PGI have developed a CUDA Fortran compiler, however the community nature of these codes means that an open technology is preferred and the proprietary nature of CUDA was a major disadvantage along with the fact that it is believed a directive based approach has significant programmability advantages over the API approach of CUDA.

\subsection{The interoperability between modern Fortran and OpenACC}
\label{interop}
While the PGI compiler supports OpenACC, the Cray compiler was selected for this work as it is generally accepted to be more mature in terms of OpenACC implementation \cite{norman2015case} and performance \cite{ghike2014directive}. The MONC model has been developed in Fortran 2003 and, as explained in section \ref{moncbg}, uses modern software engineering techniques such as a modular design. It has also been noted that components do not share data directly, but instead a common model state is passed between components and updated as required. The modularity of the code lends itself to porting aspects over to other architectures such as GPUs. Because of the enforced separation, the OpenACC directives in one component will have no impact upon other components and so it is trivial to enable or disable this functionality by simply enabling the appropriate component at runtime. The fact that components only share information via a structured model state, which a component reads from and updates, means that there is a single point of truth for specific functionality and porting can be done without worrying about side effect from other areas of the code. Whilst components have been ported to the GPU their CPU versions exist which can be used instead, therefore code optimisations necessary for good performance with OpenACC will not impact the existing CPU only code as in COSMO \cite{cosmocug}, although care needs to be taken to avoid code duplicated across components. 

This model state which is passed from one component to another is a Fortran derived (user defined) type similar to a structure in C. The model state contains numerous members, many of which are also derived types and as such has a complicated, hierarchical makeup. For instance, a number of prognostic fields are members of the model state, these fields are themselves derived types which not only contain the field data itself, but also meta-data such as whether the field is enabled, and the dimensions of the field. The OpenACC standard does not sufficiently address derived types and a conforming compiler may or may not correctly support these structures. One of the main challenges is the fact that these can reference, or point to, other memory areas on the host which must be copied across to the device recursively. The Cray compiler does provide limited deep copy support, however this is not yet mature and as such different attempts to send over the entire derived type or even members of derived types resulted in compiler errors and runtime faults. Certainly from our point of view it would have been ideal to transfer a subset of the model state derived type from the host to the device. Analysis of the compiler output also found that the Cray compiler only supports synchronous deep copy as those attempts which did compile successfully (but would not execute to completion) were silently reverting from an asynchronous to a synchronous deep copy. These issues meant that data had to be unpacked into separate, non-derived type, buffer variables and copied across from there. This requires significant boiler plate code (although no where near as much as if we had used OpenCL) and adds unnecessary verbosity bearing in mind the substantial number of fields and constants, held in the model state, required by the advection component. The need for this additional boiler plate code was unexpected and as OpenACC matures it is expected that the underlying technology will sufficiently handle these aspects, from a code point of view it will be trivial to replace this boiler plate in the future.

Two advection components are shipped with MONC, although others do exist and are likely candidates for future inclusion. The first scheme, an energy conserving scheme by Piacsek and Williams \cite{pw} is, from both a code and computation point of view, the simpler scheme of the two. The second scheme is a positivity preserving ULTIMATE scheme developed by Leonard et al \cite{tvd} and the code is far more complex and the scheme around sixteen times more computationally intensive than the PW scheme. Whilst it was our initial intention to port both of these schemes, errors and issues with OpenACC were encountered when working on the ULTIMATE scheme. The ULTIMATE scheme works by computing fluxes on columns and there is a strict dependency on the fluxes from one column being available for computation of the next. The \emph{routine vector} directives were added to indicate fine grained parallelism, however the kernels failed to launch successfully. Some discussion of the \emph{vector} directive was made in \cite{nim2014} but regardless of their work, our further investigations found an invalid device memory access was raised and whatever combination of directives or data copying used, an error pointing to memory corruption still occurs. This requires further investigation, ideally in collaboration with the compiler developers, one of the factors might be that the flux calculations we were aiming to port are over 1000 lines of code and heavily proceduralised. 

When copying the results data from the device to the host, this must be copied into a buffer present on the host and then integrated with the existing CPU source term data. This requirement for an additional host side buffer requires a significant amount of extra memory and might be problematic for applications already bound by memory. Ideally OpenACC would provide a way in which this data copy back and integration could be done in place, so that as the GPU data is being received by the host it is combined with the CPU source terms in the same memory. As our approach currently stands, the implementation of \emph{step fields} waits for all the resulting source terms to be copied back from the GPU before starting the integration. An easy way of working with incomplete data would potentially improve the performance where the source terms are integrated as they stream back, hence overlapping computation with data copying, rather than waiting for the entire data copy to complete before doing this work.

\section{Results}
We have carried out performance testing and energy usage evaluation on Piz Daint, a Cray XC30, where each node is equipped with an 8-core Sandybridge CPU, a K20X NVDIA Tesla GPU and 32GB of memory. A tank experiment test case is being performed, where a number of bubbles of air are formed throughout the system and then their movement due to being heavier (colder) or lighter (warmer) than the surrounding air is modelled. This takes into account factors such as pressure, buoyancy and the wind, along with us turning on the sub-grid scheme and five moisture (\emph{q}) variables with some simple cloud physics. The code has been compiled with the Cray compiler at optimisation level two.  

\begin{figure}
\includegraphics[scale=0.37]{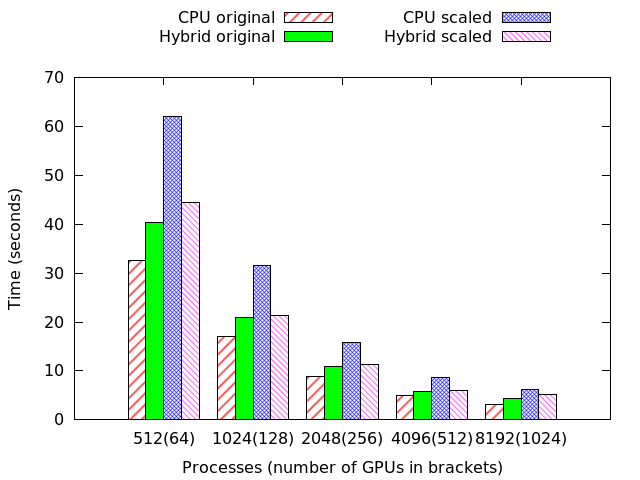}
\caption{Strong scaling, 268 million global grid points}
\label{fig:strong_scaling}
\end{figure}

Figure \ref{fig:strong_scaling} illustrates strong scaling with a fixed 3D global problem size of 268 million global grid points (z=64, x=2048, y=2048), modelling 50 seconds of simulation time. It can be seen from the graph that, for the original PW advection the CPU version is slightly more performant at all core counts than the hybrid version. This is because, as discussed in section \ref{interop}, the PW advection scheme is the least computationally intensive of the two advection schemes. The shear amount of data that needs to be transferred between host and device, even after the data transfer optimisation discussed in section \ref{imp}, outweighs any gain we get from hybridising the code and running these aspects concurrently. In this case the CPU finishes its work and the \emph{step fields} component, which integrates the CPU and GPU dynamic source terms, waits for the GPU data copying to be complete, with the CPU idle until this arrives.

In order to investigate the likely behaviour with more computationally intensive advection scheme we have scaled the workload of the PW advection scheme to match that of the ULTIMATE scheme, which is approximately sixteen times more computationally intensive. It can be seen from figure \ref{fig:strong_scaling} that these paint a different picture. This additional computation in the PW advection scheme directly increases the CPU time significantly, however the GPU times look very similar to the times recorded for the original PW advection runs. This is because the amount of data being transferred to and from the device is identical, and once the data is on the GPU then it costs very little in performance to increase the amount of computation that the GPU is performing.

\begin{figure}
\includegraphics[scale=0.37]{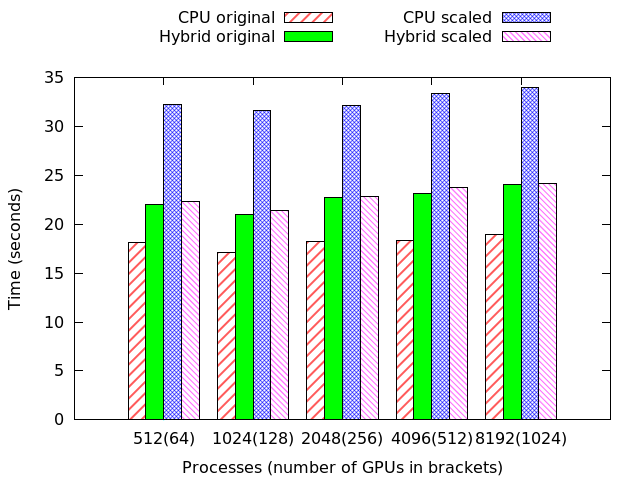}
\caption{Weak scaling, 262144 local grid points}
\label{fig:weak_scaling}
\end{figure}

Figure \ref{fig:weak_scaling} illustrates weak scaling with a fixed local problem size of 262144 grid points and the global problem size increases as we add parallelism, again modelling 50 seconds of simulation time. At 8192 processes there is a global problem size of 2.15 billion grid points. This paints a similar picture to that of the strong scaling results, where the original version favours the CPU and the scaled workload is favourable to the hybrid version. 

\subsection{Single precision}
The advection scheme discussed in this paper uses double precision, however the developers of GALES \cite{gales} previously investigated using mixed precision for their GPU code and as such the advection component was modified to use single precision. This required marshaling of the input data from double into single precision and the reserve for the resulting source terms in order to allow for the remainder of the model to run unchanged in double precision on the CPU. This is a useful optimisation strategy because, not only does single precision result in a much smaller amount of data being transferred between the host and device, it is also well known that the computation engines of the GPU are far better suited to single rather than double precision \cite{baboulin2009accelerating}. One of the major advantages of using OpenACC is that no directives or GPU specifics need to be modified, we have just changed the Fortran type of our arrays from \emph{real(kind=8)} to \emph{real(kind=4)} and the compiler takes care of everything else for the GPU to operate in single precision.  

\begin{figure}
\includegraphics[scale=0.37]{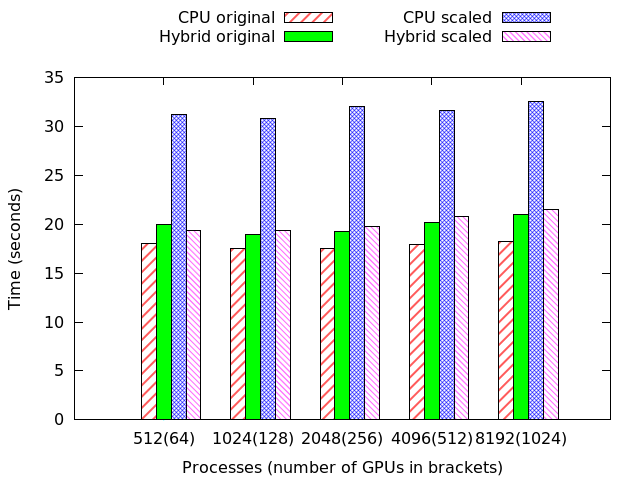}
\caption{Weak scaling single precision, 262144 local grid points}
\label{fig:weak_scaling_single}
\end{figure}

Figure \ref{fig:weak_scaling_single} illustrates weak scaling with 262144 local grid \linebreak[4]{} points on the CPU and GPU, again for the PW advection original and work scaled versions. Switching the advection scheme to single precision benefits both CPU and GPU in terms of absolute runtime numbers, on the CPU there is a slight improvement because of more efficient use of the cache. On the GPU the improvement is more significant and the gap between the original PW advection scheme on the CPU and GPU approximately halves. From the raw runtime numbers, the actual time spent by the GPU doing computation, once it has recieved the data, halves when run in single precision compared to double precision. This is consistent with published technical specifications of the K20X \cite{nvdia-datasheet}, where single precision computational performance is more than twice that of double precision performance.

\subsection{Energy usage}
\begin{figure}
\includegraphics[scale=0.37]{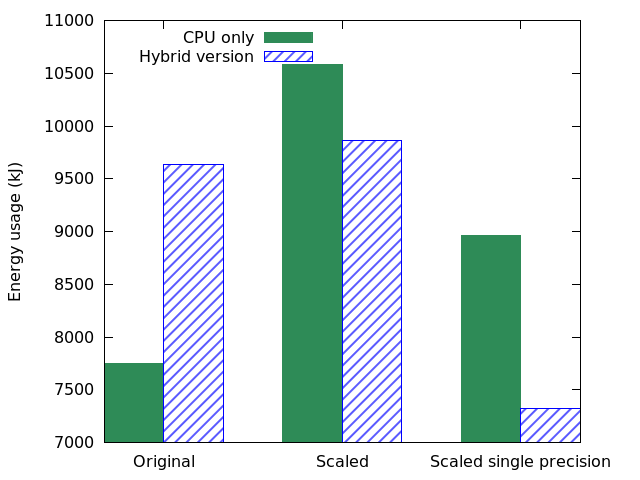}
\caption{Energy usage of different MONC configurations}
\label{fig:energy}
\end{figure}

The energy usage of codes is becoming an increasingly important consideration, not least because the community accept that in order to reach the exa-scale then current power usage will need to be cut significantly from the consumption levels of contemporary machines. It might also be the case that charging for future HPC machine use is also based upon energy usage as well as time. Figure \ref{fig:energy} illustrates the energy usage of different MONC run configurations when run over 8192 processes (1024 GPUs) for a local problem size of 262144 grid points (2.15 billion global grid points.) It is unsurprising that the basic CPU only version of this code is more energy efficient than the hybrid version, as the hybrid version takes slightly longer and runs the GPU. When the computation is scaled up the energy usage results are surprising. The hybrid version is more energy efficient than the CPU version which we were not expecting because, even though the hybrid scaled version completes far sooner than the CPU only version, GPUs are well known to have high energy requirements. The single precision, computation scaled energy usage was then compared and from this it is clear that single precision improves the energy efficiency of both versions. This is a combination of the fact that the model is both completing faster and, in the case of the hybrid run, less data is being transferred to and from the device. This energy usage data was collected via Cray's Resource Utilization Reporting (RUR) system \cite{rurcug} which instructs a node's operating system to track the energy usage. The local level energy usage from each node is then combined by RUR at the end of a job run into an overall total which is written to a file in the user's home directory.  

\section{Conclusions and further work}
In this paper we have described the use of directives, via OpenACC, to develop a hybrid LES model for climate and weather research. We have concentrated on off-loading the most computationally intensive single part of the model onto the GPU, whilst running the rest of the dynamical core on the CPU concurrently before integrating the results together. We have demonstrated that with increased computational workloads, as found in more advanced advection schemes, then this approach has the potential to improve overall model runtime significantly. Whilst OpenACC is a reasonable choice for use with both modern Fortran and this sort of code acceleration, we have highlighted some shortcomings of the technology both from a feature and maturity point of view. We have demonstrated that by switching to single precision one can gain improvements in hybrid runtime and that OpenACC provides the flexibility to easy accomplish this. Our energy experimentation has illustrated that if one aims for performance then it is likely that this will have a side effect of also maximising energy efficiency. Whilst there is an energy penalty when the GPU is underutilised compared with a straightforward CPU code, once a significant amount of computation is being performed by the GPU then the overall saving in runtime becomes more significant.

An important aspect of further work will be that of auto tuning. Performance of the hybrid code is optimal when the CPU and GPU have roughly equal workload. Based upon the work described in this paper, other aspects of the dynamical core would be fairly trivial to port via OpenACC. Because these components are enabled or disabled at runtime, by making available both the CPU and GPU versions of these activities then the model could analyse the performance of various hybrid permutations over a number of timesteps and then select which components to run where in order to achieve an optimally balanced load. Another important consideration is that these dynamical components all operate on the same prognostic fields, so very little or no additional data would need to be copied across other than that which is already transferred for the advection scheme. Equally, on copying the data back from the device to the host, the GPU could combine all the source terms for the components it has been running, before the copy back is performed so the same amount of data is transferred back to the host regardless of the number of dynamical core components running on the GPU. Taking advantage of these data properties by running multiple different kernels concurrently would make more efficient use of the accelerator. 

Whilst we still believe that OpenACC has been a good fit for this work, the technology and compiler support has some way to go to reach full maturity. Once it does then this technology will fit  well with modern Fortran and software engineering techniques such as the modularity employed by MONC. A more mature ecosystem, from the implementation level to documentation, will support scientists from a variety of disciplines adding directives to well engineered existing code in order to enable it to take advantage of these heterogeneous architectures. More work is to be done investigating the issues surrounding the use of OpenACC with the more complex ULTIMATE advection scheme, it is our belief that this will require coordination from the compiler developers to understand what is happening here and the correct course of action to fix it. Due to the performance illustrated in this paper when running larger work loads on the GPU, we believe that this is activity is an important one in order to make most efficient use of the technology.

In terms of weather and climate models, the science and code size of MONC model is relatively simple when compared to other models such as the Met Office's Unified Model. However it is believed that the techniques and lessons learnt from this work can equally be applied to other, more complex, models. Using directives, either OpenACC or OpenMP 4.0 when it reaches maturity, to offload the computationally intensive aspects of these codes and to run them in a hybrid fashion has huge potential.

\section{Acknowledgments}
The authors would like to express their gratitude to the Swiss National Supercomputing Centre, CSCS, who have kindly given us access to and time on Piz Daint for this work. 
\bibliographystyle{plain}
\bibliography{sigproc}
\end{document}